\documentstyle[aps,preprint]{revtex}

\begin{document}
\title{Electric Field Induced Phase Transition in KDP Crystal Near Curie
Point: Raman and X-ray Scattering Studies}
\author{A. T. Varela\footnote{Permanent address: Centro Federal de %
Educa\c{c}\~{a}o Tecnol\'{o}gica do Cear\'{a}}, J. M. Sasaki, I.
Guedes, P. T. C. Freire, A. P. Ayala , J. Mendes Filho, F. E. A.
Melo\footnote{e-mail: erivan@fisica.ufc.br}, and A. S.
Chaves\footnote{Permanent address: Universidade de Bras\'{\i}lia,
Bras\'{\i}lia-DF, Brazil}}
\address{Departamento de F\'{\i}sica, Universidade Federal do Cear\'{a}, %
Campus do Pici, P.O.Box 6030, 60455-760 Fortaleza, CE, Brazil}
\maketitle

\begin{abstract}

X-ray scattering measurements are performed in order to verify %
that the mechanism leading to the DC electric field induced
$C_{2v}^{19} \rightarrow C_{2v}^{ \neq 19}$ phase transition in
KDP crystal at 119 K is the changing of the local sites symmetries
of phosphate group from $C_2$ in the $C_{2v}^{19}$ phase to $C_s$
in the $C_{2v}^{ \neq 19}$ phase. It is shown by analyzing the
integrated intensity of the (800) and (080) reflections that under
DC electric field  the density of oxygen atoms lying on these
plane changes indicating that phosphate group rotates around the
[010] direction relative to the orthorhombic $C_{2v}^{19}$
structure. Some Raman results are also discussed.
\end{abstract}

\newpage

Among the several ferroelectric materials which contains hydrogen
bonds, potassium dihydrogen phosphate (KDP) is probably the most
investigated. At a temperature of 122 K, the KDP crystals undergo
a ferroelectric phase transition , where occurs the lowering of
the crystal symmetry from the tetragonal $D_{2d}^{12}$ phase to
the orthorhombic $C_{2v}^{19}$ phase. As a result the crystal
lattice becomes polarized along the $c-axis$. Near and below the
phase-transition temperature the protons are partially ordered
[1], being located near either the upper or the lower oxygen atoms
of phosphate groups . Many works reporting on the investigation of
the stability of both $D_{2d}^{12}$ and $C_{2v}^{19}$ phases as a
function of the hydrostatic pressure and low-intensity  DC
electric fields were already published [2-7]. Recently, we have
investigated the effect of uniaxial pressure on these KDP phases
[8]. We have shown that under uniaxial pressure, where the force
was applied along the shear direction, the KDP undergoes two
metastable transitions, namely: {\it (i)} $D_{2d}^{12} \rightarrow
C_{2v}^{j \neq 19}$ and {\it (ii)} $C_{2v}^{19} \rightarrow
C_{2v}^{j \neq 19}$. A reasonable explanation for these
transitions was given based on the changing of the local site
symmetry of the phosphate ions. In the ferroelectric phase, the
phosphate ion changes its local site symmetry from $C_2$ to $C_s$
, maintaining the same factor group $C_{2v}$ but modifying the
space group. These changes are due to the rotation  of  the
phosphate ions around the [010] direction of the orthorhombic
structure. (It is important to mention  that under an electric
field and low temperature a phase transition from a monoclinic
$C_s$ structure to an orthorhombic $C_{2v}$ one was observed [9]).
Also, after discussing the reversibility criteria of this new
metastable phase, we have drawn the phase diagram for the KDP
transitions on the plane $(\sigma_6,T)$ for temperatures in the
range from 110 K to 130 K.  A theoretical explanation to the
appearance of the metastable $C_{2v}^{j \neq 19}$ phase based on
the Gibbs free energy density of the system was given. We have
expanded the phenomenological Gibbs free energy density of system
up to $P_3^{10}$, where $P_3$ is the spontaneous polarization
along the [001] direction. Even  for small values of the
coefficient of the term $P_3^{10}$, a second minimum for the
crystal energy can be achieved. This second minimum is associated
to the metastable $C_{2v}^{j \neq 19}$ phase, which presents a
lower value of polarization than that presented by the stable
$C_{2v}^{19}$ phase. This is in accordance with our assumption
that the dipoles rotate around the orthorhombic $b-axis$ when KDP
undergoes a phase transition. This assumption can also be verified
using other experimental techniques, e.g., X-ray diffraction,
where we can observe modifications in the diffraction pattern
associated with the $(h00)$ and $(0k0)$ reflection planes, with
$h,k = 4n$ where $n = 1,2,...,$ once the number of oxygen atoms
lying on these planes change. However, it is somewhat difficult to
perform X-ray scattering measurements using the stress apparatus
described in Ref.[8]. So, we decided take advantage of the fact
that the ferroelectric phase present piezoelectricity, to
investigate the phase transition $C_{2v}^{19} \rightarrow
C_{2v}^{j \neq 19}$ under DC electric field applied along the
[001] ferroelectric direction. In other words, due to the converse
piezoelectric effect a DC electric field applied along the [001]
direction should induce phase transitions in KDP crystals
similarly to that induced by uniaxial pressure with the force
applied along the shear direction. Hence, the main goal of this
work is to perform x-ray diffraction measurements to verify the
assumption that due to a DC electric field applied along the
ferroelectric $c-axis$, the dipoles rotate around the [010]
direction of the orthorhombic structure leading to a change in the
local site symmetry of the phosphate ions.

The samples were cut from good optical quality crystals grown by
slow evaporation in parallelepipeds of dimensions  $6 \times 5
\times 1.5$ mm$^3$ and oriented by X-ray diffraction .  The
parallelepipeds faces were orthogonal to the [100], [010] and
[001] directions to the orthorhombic  structure. Electrodes of
silver were evaporated on the large faces which are perpendicular
to the ferroelectric [001] direction. A Keithley Instruments
voltage supply model 246 was used as the voltage source.

Light scattering measurements were performed using a conventional
equipment  (argon ion laser and double monochromator), with an
experimental resolution of 1 cm$^{-1}$. A continuos flow-type
cryostat was used  to record the Raman spectra at low
temperatures, which could be controlled to $\pm$ 0.1 K. Geometries
for the spectra listed in the figures follow the usual Porto
notation A(BC)D. X-ray scattering measurements were performed
using a Rigaku diffractometer with radiation source of Mo
K$\alpha$ coupled with a low temperature chamber. The good
penetration of X-ray beam gives advantage to get diffraction from
deeper planes, avoiding the non uniformity of electric field on
surface. The procedure used in the X-ray experiments was: first,
the crystal was aligned using the (440) reflection of paraelectric
phase and then cooled down to ferroelectric phase where during the
transition, the reciprocal lattice points of  (800) and (080)
reflections appear satisfying the diffraction conditions of the
orthorhombic structure with the $C_{2v}$ factor group.

Before discussing X-ray results, we need to show that under DC
electric field, KDP crystal undergoes the $C_{2v}^{19} \rightarrow
C_{2v}^{j \neq 19}$  phase transition. Then, in Fig. 1(a) and 1(b)
we show part of DC electric field dependent Raman spectra for the
low frequency taken at 119 K, for the symmetries $A_1$ and $B_1$
of the $C_{2v}$ factor group, respectively.  For $E = 0$, both
$A_1$ and $B_1$ spectra present a good agreement with the mode
distribution predicted by the group theory analysis. However, for
$E = 5$ kV/cm , we observe qualitative modifications in the Raman
spectra. From Fig. 1(a) we can see three modifications: {\it (i)}
The increase of the vibration for $\omega < 250$ cm$^{-1}$; {\it
(ii)} the increase in the intensity of the vibration at 525
cm$^{-1}$ and {\it (iii)} the disappearance of the vibration at
575 cm$^{-1}$. For $B_1$ symmetry, as shown in Fig. 1(b), we
observe an inversion in the intensity of the peaks oscillating at
around 200 and 500 cm$^{-1}$. Since the symmetries $A_1$ and $B_1$
are unidimensional , the modifications exhibit by the Raman
spectra are an evidence that due to the DC electric field the
$C_{2v}^{19}$ phase do KDP underwent a phase transition .The
modifications observed comply with the symmetry analysis
considering the transition from the space group $C_{2v}^{19}$ to
the space group $C_{2v}^{ \neq 19}$, where the phosphate ion
changes its local site symmetry from $C_{2}$ to $C_{s}$.  The
modifications observed are irreversible once all features seen in
the Raman spectra of Figs. 1(a) and 1(b) remain present even when
the DC electric field is turned off, and the crystal is maintained
in this condition for an arbitrary long time. To go back to the
spectra of KDP for the $C_{2v}^{19}$ phase, we must increase the
temperature of the crystal over 122 K, and then cool it again at
temperatures below 122 K. This irreversibility can be understood
as a manifestation of a lowering in the cell potential due to an
increase in the dipole interactions. Switching off the DC electric
field is not sufficient to overcome the potential barrier created
by  dipolar interaction. This barrier is overcame only by
transferring thermal energy to the dipoles. Due to these facts, we
conclude that a DC electric field  induces a phase transition in
KDP similar to that induced by uniaxial pressure.

In order to present another experimental evidence of the mechanism
leading to the appearance of the $C_{2v}^{j \neq 19}$ phase, we
have performed single crystal X-ray measurements as a function of
DC electric field. The idea of the experiment is very simple: if
the phosphate tetrahedron in fact rotates around the [010]
direction, a variation in  the behavior of the integrated
intensity of diffraction peaks associated with (800) and (080)
reflection planes should be observed, once the number of oxygen
lying on these planes changes when E is varied from 0 up to 5
kV/cm, as shown in  Fig. 2. Figure 3 shows the diffraction
patterns corresponding to the (800) and (080) reflection planes of
the orthorhombic structure as a function of the DC electric field
up to 5 kV/cm at 119 K. The features observed are irreversible in
the same way that those presented in the Raman spectra. The peaks
emerge from overlapping of the bands corresponding to the $K
\alpha_1$ and $K \alpha_2$ lines of  Mo radiation. By performing a
spectral decomposition into pseudo-Voigth  components, we can draw
the behavior of the integrated intensity of the (800) and (080) $K
\alpha_{1}$ reflections as a function of DC electric field as
displayed in Fig.4. It should be observed that the integrated
intensity corresponding to (800) reflection decreases with
increasing the DC electric field up to 5 kV/cm, while that one of
(080) reflection increases. These changes in the integrated
intensity behavior indicate that the density of oxygen atoms lying
on these planes changed. This can be ascribed as a result of a
rotation of phosphate ion around the [010] direction relative to
the $C_{2v}^{19}$ orthorhombic structure. This statement agrees
with the observation performed by Bacon and Pease [10], where they
showed  that the KDP under  DC electric field exhibits the value
of the saturation polarization of the order of $4.7 \times
10^{-6}$ Ccm$^{-2}$, whereas the observed value is $5 \times
10^{-6}$ Ccm$^{-2}$ for $E = 0$ [11] . Due to this fact, the local
sites symmetries exhibited by the phosphate ion changes  from
$C_{2}$ to $C_{s}$ which modifies the space group of the $C_{2v}$
symmetry from $j=19$ to $j \neq 19$. In conclusion we reported on
the experimental verification based on X-ray measurements of the
mechanism leading to the conformational $C_{2v}^{19} \rightarrow
C_{2v}^{ \neq 19}$ phase transition of KDP when a DC electric
field is applied along the ferroelectric c axis. The behavior of
the integrated intensity of (800) and (080) reflections indicates
that occurs a modification in the density of oxygen atoms lying on
these planes . This modification results from the rotation of
phosphate ion around the orthorhombic [010] axis. This rotation
changes the local sites symmetries of phosphate ion from $C_{2}$
in the $C_{2v}^{19}$ phase to $C_{s}$ in the $C_{2v}^{ \neq 19}$
phase.

\begin{center}
{\bf ACKNOWLEDGEMENTS}
\end{center}

Financial support from CAPES, CNPq, FINEP and FUNCAP, Brazilian
funding agencies, is gratefully acknowledged.

\begin{center}
{\bf REFERENCES}
\end{center}

\parindent0.0truecm
[1] R.J. Nelmes, W.F. Kuhs, C.J. Howard, J.E. Tibballs and T.W.
Ryan, J. Phys. C: Solid State Phys. 18, L711 (1985).

[2] G. Busch, Helv.Phys.Acta 11,269 (1938).

[3] F. Jona and G. Shirane, Ferroelectric Crystals (Dover, New
York, 1993).

[4] K.Itoh,T. Matsubayashi,E. Nakamura, and H. Motegi,
J.Phys.Soc.Jpn. 39, 843 (1975).

[5] F.E.A.Melo, K.C.Serra, R.C.Souza, S.G.C.Moreira,
J.Mendes-Filho, and J.E.Moreira, Braz. J. Phys. 22,95 (1992).

[6] B.Morosin and G.Samara, Ferroelectrics 3, 49 (1971).

[7] P.S.Pearcy and G.Samara, Phys. Rev.B 8, 2033 (1973).

[8] F.E.A. Melo, S.G.C. Moreira, A.S. Chaves, I. Guedes, P.T.C.
Freire, and J. Mendes Filho, Phys. Rev. B 59, 3276 (1999).

[9] S.G.C. Moreira, F.E.A. Melo, and J. Mendes Filho, Phys. Rev. B
54, 6027 (1996).

[10] G. E. Bacon and R.S. Pease,  Proc. R. Soc. A 230, 359 (1955).

[11] A. von Arx and W. Bantle, Helv. Phys. Acta. 16, 211 (1943).

\newpage

\begin{center}
{\bf FIGURE CAPTIONS}
\end{center}

Fig. 1- Raman spectra of KDP as a function of the DC electric
field at 119 K for two different symmetries: (a) $A_1$ and (b)
$B_1$.

Fig. 2- Schematic representation of the orthorhombic $C_{2v}^{19}$
structure of the KDP projected on the ab plane.

Fig. 3- Experimental single crystal X-ray diffraction pattern
related to (800) and (080) reflection planes as a function of the
DC electric field at 119 K.

Fig. 4- Plots of the integrated peak intensity corresponding to
(800) and (080) $K \alpha_{1}$ reflections as a function of the DC
electric field at 119 K.

\end{document}